\documentstyle[12pt,epsf]{article}\textheight 230mm\textwidth 160mm
\newcommand{\bi}{\bibitem}
\newcommand{\be}{\begin{eqnarray}}
\newcommand{\ee}{\end{eqnarray}}
\newcommand{\nn}{\nonumber}
\catcode`\@=11
\def\lsim{\mathrel{\mathpalette\@versim<}}
\def\gsim{\mathrel{\mathpalette\@versim>}}
\def\@versim#1#2{\vcenter{\offinterlineskip
\ialign{$\m@th#1\hfil##\hfil$\crcr#2\crcr\sim\crcr } }}
\catcode`\@=12

\begin{document}
\pagestyle{empty}
\hspace*{9cm}\vspace{-3mm}MPI-PhT/95-133, DFTT 80/95\\ 
\hspace*{9.7cm}\vspace{-3mm}KANAZAWA-95-20\\
\begin{center}
{\Large\bf  Testing Gauge-Yukawa-Unified Models
By $M_t$ }
\end{center} 

\vspace{1cm}

\begin{center}{\sc Jisuke Kubo}$\ ^{(1),*}$, 
{\sc Myriam
Mondrag{\' o}n}$\ ^{(2),**}$, \vspace{-1mm}\\
{\sc Marek Olechowski}$\ ^{(3),\dag}$ and 
{\sc George Zoupanos}$\ ^{(4),\ddag}$  
\end{center}
\begin{center}
{\em $\ ^{(1)}$ 
 College of Liberal Arts, Kanazawa \vspace{-2mm} University,
Kanazawa 920-11, Japan } \\
{\em $\ ^{(2)}$ Institut f{\" u}r Theoretische Physik,
Philosophenweg 16 \vspace{-2mm}\\
D-69120 Heidelberg, Germany} \vspace{-2mm}\\
{\em $\ ^{(3)}$ INFN Sezione di Torino and 
Dipartamento di Fisica Teorica, Universit{\` a} 
di Torino \vspace{-2mm}\\
      Via P. Giuria 1, 10125 Turin, Italy}\vspace{-2mm}\\
{\em $\ ^{(4)}$ Max-Planck-Institut f\"ur Physik,
 Werner-Heisenberg-Institut \vspace{-2mm}\\
D-80805 Munich, Germany}   \end{center}

\begin{center}
{\sc\large Abstract}
\end{center}

\noindent
Gauge-Yukawa Unification (GYU) relates \vspace{-3mm} the gauge and
Yukawa couplings, thereby
 going beyond the
usual GUTs, \vspace{-3mm} and it is assumed that
the GYU in the third fermion generation
implies that its Yukawa couplings are \vspace{-3mm}
of the same order as the unified gauge coupling
at the GUT scale.  We
 re-examine  carefully the recent \vspace{-3mm} observation that 
the top-bottom mass hierarchy can be explained 
\vspace{-3mm} to a certain
extent  in supersymmetric GYU models.  It is found that
there are equiv-top-mass-lines 
in the boundary conditions of  \vspace{-3mm} the Yukawa
couplings so that two different GYU
models on the same line can \vspace{-3mm}
not be distinguished by the top mass $M_t$ alone. 
If they are  on different lines, \vspace{-3mm} they could be 
distinguished by $M_t$ in principle, provided
that the predicted $M_t$'s are \vspace{-3mm} well below the infrared
value  $M_t$ (IR). 
We find that the ratio $M_t ({\rm IR})/\sin\beta$  \vspace{-3mm}
depends 
on $\tan\beta$ for large $\tan\beta$ and  the lowest
value  of $M_t ({\rm IR})$ is $\sim 188$ GeV.
 \vspace{-3mm} We focus our
attention on the existing $SU(5)$ 
GYU models which are obtained by \vspace{-3mm} requiring finiteness
and reduction of couplings.
They, respectively, predict  
$M_t= (183+\delta^{\rm MSSM} M_t\pm 5)$  GeV and \vspace{-3mm}
$(181+\delta^{\rm MSSM} M_t\pm 3)$ GeV, 
where $\delta^{\rm MSSM} M_t$ stands 
for the MSSM threshold correction and is $\sim -2$ GeV 
 \vspace{-3mm} for the case
that all the MSSM superpartners have the same mass $M_{\rm SUSY}$
with $\mu_H/M_{\rm SUSY} <<1$.

\vspace*{1.5cm}
\footnoterule
\vspace*{2mm}
\noindent
$^{*}$Partially supported  by the Grants-in-Aid
for \vspace{-3mm} Scientific Research  from the Ministry of
Education, Science 
and Culture \vspace{-3mm}  (No. 40211213).\\
\noindent
$^{**}$Address  after Jan. 1st, '96:
Instituto de F{\' \i}sica,  \vspace{-3mm} UNAM,
Apdo. Postal 20-364,
M{\' e}xico 01000 D.F., \vspace{-3mm} M{\' e}xico.\\
\noindent
$ ^{\dag}$On leave of absence  from
the Institute of \vspace{-3mm} Theoretical Physics,
Warsaw University,
ul. Hoza 69, 00-681 Warsaw, \vspace{-3mm} Poland.
Partially supported  by the Polish Committee for Scientific
Research. \vspace{-3mm}\\
\noindent
$ ^{\ddag}$On leave \vspace{-3mm} of absence from 
Physics Department, National
Technical University,  GR-157 80 Zografou, Athens, Greece.  
Partially supported \vspace{-3mm} by C.E.C. projects, SC1-CT91-0729;
CHRX-CT93-0319.

\newpage
\pagestyle{plain}

\section{Introduction}

The great success of the standard model (SM) 
of the electroweak and strong interactions 
is spoiled by the presence of the plethora of its free parameters.  The
traditional way to reduce the number of independent parameters in  a
theory is to introduce a symmetry. 
Grand Unified Theories (GUTs) \cite{pati1,georgi2,fritzsch1}
 are representative examples
of such attempts.
The SU(5) GUTs, for instance,  reduce by one the gauge couplings of
the SM and provide us with the  prediction for
one of them \cite{gqw}. GUTs can also relate the Yukawa couplings among
themselves, and in turn might lead to testable predictions for the SM
parameters. The prediction of the mass ratio $M_{\tau} / M_b$  in
the minimal SU(5)
was a successful example of reduction of the independent
parameters of the Yukawa sector \cite{buras1}.

In general, the gauge and Yukawa sectors in GUTs are not related.
In searching for a symmetry which could relate the two 
sectors, one is
naturally led to introduce supersymmetry, given that the fields
involved have different spins.
It, however, turns out that one has to introduce at least N=2
supersymmetry to 
understand the Yukawa sector as a part of the gauge 
sector \cite{fayet1}.
This is a very strong constraint for the construction of
realistic theories  \cite{aguila1},
because the models based on extended supersymmetries do 
not possess a chiral structure.
In superstring and composite models,  
such relations  exist also. 
But in spite of some recent developments, 
 there exist open problems which  are partly related to 
the lack of realistic models.

By a Gauge-Yukawa Unification (GYU) we mean a functional
relationship among the gauge and Yukawa couplings, which
can be derived from some principle. 
In contrast to the above mentioned schemes, in the GYU scheme based on the
principle of reduction of couplings 
\cite{zimmermann1}--\cite{kubo33} and
finiteness \cite{finite1}--\cite{mondragon1}, one can write down
relations among the gauge and Yukawa couplings
 in a more concrete fashion.
These principles, which are formulated 
within the framework of 
perturbatively renormalizable field theory, are not explicit symmetry
principles, although they might imply symmetries.
 The former principle is based on the fact that
there may exist renormalization group (RG)
invariant relations among couplings which preserve
 perturbative
renormalizability \cite{zimmermann1}. And the latter one is based 
on the possibility that these RG
invariant relations among couplings lead to
 finiteness in perturbation theory \cite{finite1,finite2},
even to all orders \cite{finite4,sibold1}.
Theoretical possibilities of relating
couplings discussed here exhibit a
generalization of the traditional renormalizability:
One can reduce the number of  independent 
couplings without introducing necessarily a symmetry,
thereby improving the
calculability and predictive power of a given theory
and hence 
generalizing the notion of naturality \cite{nat} in a certain sense.

The consequence of a GYU is that 
in the lowest order in perturbation theory
the gauge and Yukawa couplings  are related in the form
\be
g_i& = &\kappa_i \,g_{\rm GUT}~,~i=1,2,3,e,\ldots,\tau,b,t
\ee
above the unification scale $M_{\rm GUT}$,
 where $g_i~(i=1,\ldots,t)$ stand
for the gauge  and Yukawa couplings, and $g_{\rm GUT}$ is the unified
coupling. (We have neglected  the Cabibbo-Kobayashi-Maskawa mixing 
of the quarks.)
 The constants $\kappa$'s can be explicitly
calculated in the GYU scheme based on the principle of reduction of
couplings, and it has been found
\cite{kubo2}--\cite{kubo33}, \cite{mondragon1} that various 
supersymmetric GYU  models can predict $M_t$ and $M_b$
that
are consistent with the present experimental data \cite{pdg,top,hagiwara1}. 
This means that the top-bottom hierarchy 
could be
explained to a certain extent in these models in which 
one assumes the existence of a GYU at the unification scale
$M_{\rm GUT}$,
which should be compared with how
the hierarchy of the gauge couplings
can be explained if one assumes  the existence of a unifying
gauge symmetry at $M_{\rm GUT}$ \cite{gqw}.

It has been observed \cite{kubo2}--\cite{kubo33}, \cite{mondragon1} 
that there exists a
relatively wide range of $\kappa$'s of $O(1)$ which gives the 
top-bottom hierarchy of the right order. 
Of course, the existence of this range is partially related to the
infrared behavior of the Yukawa couplings \cite{hill1,bardeen}.
However, because of the restricted number of analyses which have been
performed so far, it was not possible to conclude whether the
calculated $\kappa 's$ in each case give predictions for
$M_{t}$ and $M_{b}$, which are consistent with
the experimental data just because the large
experimental as well as theoretical uncertainties, 
or whether the top-bottom hierarchy results mainly
from the infrared behavior of the Yukawa couplings, and therefore the
precise nature of a GYU 
is not important.
It is therefore crucial,
in order to test
GYU models by
more precise measurements of $M_{t}$ as well as $M_b$,
 to see
which of the cases is indeed  realized.
More precise and systematic investigations
on the range of $\kappa$'s are moreover
indispensable
in constructing realistic GYU  models 
and in distinguishing them from each other by experiments.

For this purpose we fist calculate the infrared 
quasi--fixed--point
value of $M_t$ \cite{hill1,bardeen}
(which we denote by $M_t$(IR))
for large $\tan\beta$.
We find that the $\tan\beta$ dependence of
the ratio $M_t ({\rm IR})/\sin\beta$ 
 for large $\tan\beta$ ($\gsim 40$) is not negligible and the lowest
value  of $M_t ({\rm IR})$ is $\simeq 188$ GeV.
We also find that
there exist equiv-top-mass lines 
in the space of the boundary conditions of  the Yukawa
couplings so that two different GYU
models on the same line can 
not be distinguished by $M_t$. 
(The predictions on other parameters such as $M_b$ 
varies of course along this line.)

One of our main results is
that  the present experimental data on $M_t$ and $M_b$ might
be interpreted as indicating GYU and
different supersymmetric GYU
 models could be  distinguished and tested by
a precise measurement of $M_t$ with an uncertainty of
 few GeV, provided
that the models are not on equiv-$M_t$ lines that are very 
close to each other and
the predicted $M_t$'s are well separated
from   the infrared value.
Using the updated experimental data on the SM parameters,  
we  re-examine the $M_t$ prediction of two
existing $SU(5)$ GYU models, Finite
Unified Theory based  \cite{mondragon1} and the 
asymptotically-free minimal
supersymmetric GUT with the GYU in the
 third generation \cite{kubo2}.
They predict  $M_t= (183+\delta^{\rm MSSM} M_t\pm 5)$ GeV and
$(181+\delta^{\rm MSSM} M_t\pm 3)$ GeV, respectively, 
where $\delta^{\rm MSSM} M_t$ stands  for the MSSM threshold
correction.
 We find it is $\sim -1$ \% for the case that
all the superpartners have the same mass $M_{\rm SUSUY}$
and $\mu_H/M_{\rm SUSUY} \ll 1$.

\section{The gross behavior of the Yukawa couplings}
Before we come to more complete analysis that 
among other things includes
two-loop effects, let us 
investigate within the one-loop 
approximation how the low energy values of
the Yukawa couplings $g_t,g_b$ and $g_{\tau}$ depend on
the GYU boundary 
condition (1).
Since the qualitative  behavior of the Yukawa couplings
 for the energy range relevant
to our problem can be understood without $g_1$ and $g_2 $, 
we neglect them. 

To begin with, we 
eliminate the $\mu$-dependence of the couplings
through $(\mu d/d \mu) \alpha_3 =-(3/2\pi) \alpha_{3}^{2}$
to obtain
\be
-3\,\alpha_3\frac{d \rho_{t}}{d\alpha_3} &=&\rho_{t}\,(\,
6\,\rho_{t}+\rho_{b}-\frac{7}{3}\,)~,\nn\\
-3\,\alpha_3\frac{d \rho_{b}}{d\alpha_3} &=&\rho_{b}\,(\,
\rho_{t}+6\,\rho_{b}
+\rho_{b}\,\rho_{r}-\frac{7}{3}\,)~,\\
-3\,\alpha_3\frac{d \rho_{r}}{d\alpha_3} &=&\rho_{r}\,(\,
-\rho_t-3\rho_{b}
+3\,\rho_{r}\,\rho_{b}+\frac{16}{3}\,)~,\nn
\ee
where 
\be
\alpha_3 = |g_{3}|^{2}/4 \pi~,~ \alpha_i = |g_{i}|^{2}/4 \pi~,~
\rho_i &=&\frac{\alpha_i}{\alpha_3}~,~i=t,b,\tau~,~
\rho_r =\frac{\alpha_{\tau}}{\alpha_b}~.
\ee
Then we assume a GYU so that the $\rho_i$'s 
may be assumed to be
of $O(1)$ at $M_{\rm GUT}$.
The terms containing $\rho_i$'s in the parenthesis 
in the evolution equation for $\rho_r$ are
small compared with $16/3$ if 
the $\rho_i$'s do not increase very much as $\alpha_3$ varies
from $\alpha_{\rm GUT}$ to $\alpha_3(M_{\rm SUSY})$.
Neglecting these terms further, we obtain
$\rho_{r}\simeq (\kappa_{\tau} /\kappa_{b})^2
(\alpha_3/\alpha_{\rm GUT})^{-16/9}$ which,
with $  \kappa_b /\kappa_{\tau}
\sim O(1)$ and $\alpha_3/\alpha_{\rm GUT}
\simeq 2.7$ for $M_{\rm SUSY} \sim M_Z$,
is about $0.17$. We therefore neglect
$\rho_r$ in the evolution of $\rho_b$ further so that 
the evolutions of $\rho_b$ and $\rho_t$ become
symmetric. We then find that, if 
$\rho_b /\rho_t \sim O(1)$ at $M_{\rm GUT}$,
the ratio  at low energies roughly remains the same. So we assume
that the solution of
\be
-3\,\alpha_3\frac{d \rho_{t}}{d\alpha_3} &=&\rho_{t}\,(\,
7\,\rho_{t}-\frac{7}{3}\,)~,\nn\\
-3\,\alpha_3\frac{d \rho_{r}}{d\alpha_3} &=&\rho_{r}\,(\,
-\gamma \rho_t+\frac{16}{3}\,)
\ee
can describe the gross behavior of $\rho_i$'s,
where we have introduced $\gamma (< 4)$ to take into
account approximately the $\rho_{r} \rho_b $-term in the
evolution of $\rho_r$.  We find that the solution is given by
\be
\rho_{t}(M_{\rm SUSY}) &\simeq&
\frac{1}{3 +(\kappa_{t}^{-2}-
3)[\alpha_{\rm GUT} /\alpha_3 (M_{\rm SUSY})]^{7/9}}~,\nn\\
\rho_r (M_{\rm SUSY}) &\simeq& 
\kappa_{r}^{2}\,
[\frac{\kappa^{2}_{t}}{\rho_{t}(M_{\rm SUSY})}]^{\gamma/ 7}\,
[\frac{\alpha_{\rm GUT}}{\alpha_{3}(M_{\rm SUSY})}]^{(16-\gamma)/9}~,
\ee
where $\kappa_{t~(r)}=\rho_{t~(r)}(M_{\rm GUT})$.
That the factor $ [\alpha_{\rm GUT} /\alpha_3]^{7/9}$
goes to zero as $ \alpha_3 $ approaches $\infty$ comes from the
Pendleton-Ross
infrared-fixed-point behavior of $ \rho_t$ \cite{pr}.
For the present case, it is about $ (1/2.7)^{7/9} \simeq 0.46  $
 so that 
the low energy value of $\rho_t$ can not be explained solely 
from this fixed point behavior,
except for the case that the $\kappa_{t}^{2}$ is very close to $1/3$. 
But this factor is small so that 
the $\rho_{t}(M_{\rm SUSY})$ depends  weakly
on $\kappa_{t}^{2} $, and especially for  large $\kappa_{t}^{2}$
this dependence disappears practically, which is
Hill's observation of the intermediate-fixed-point
\cite{hill1,bardeen}. 

It is, therefore, crucial for the testability of
GYU models by the $M_t$ prediction that
$M_t$ is sufficiently different from the infrared value.
Of course, how much $M_t$ should be away from the
infrared value depends on the
experimental accuracy.
In the next section, we will discuss
this problem more in detail.
Within the present approximation we may 
conclude that 
\be
|\frac{\Delta \rho_t}{\rho_t}| &\simeq & 
(\,0.92-0.22 \,)\,|\frac{\Delta \kappa_t}{\kappa_t}|~~
\mbox{for}~~\kappa_t \simeq 0.5 - 1.5~.
\ee
Since $M_t \propto \sqrt{\rho_t}$, an uncertainty of
$2$ \% in $M_t$ for instance will allow
the range of $\kappa_t$ that corresponds to
$| \Delta \kappa_t/\kappa_t|
\simeq 0.04 - 0.18$. If the uncertainty is of $O(10 \%)$,
one finds that there is a wide range of the 
allowed values of $\kappa_{t}^{2}$, which qualitatively explain the
observation of refs. \cite{kubo2}--\cite{kubo33}\cite{mondragon1}.

It should be stressed that, to calculate
the fermion masses, we need  to 
know the value of  $\tan \beta$
\cite{inoue1} in addition to
 the Yukawa couplings, which should be
 contrasted to the case of the SM.
At the tree level, it can be expressed as
\be
\tan \beta &=& [\,2\, (\frac{M_{W}^{2}}{M_{\tau}^{2}})\,
\frac{\alpha_3}{\alpha_2}\,\rho_{r}\rho_b-1\,]^{1/2}\nn\\
&\simeq&  111\sqrt{\rho_r \rho_b}~\simeq ~60.8\,\sqrt{\rho_t}
~~\mbox{for}~~\rho_r \simeq 0.3 ~~\mbox{and}~
~\rho_b \simeq \rho_t~.
\ee
>From eq. (7), we see that $\tan\beta$ can be predicted
from a GYU if we use $\sin^{2}\theta_{W}, \alpha_{\rm EM},
M_Z$ and $M_{\tau}$ as inputs and it will be large for
GYU models \footnote{This result was previously obtained in
a different context in ref. \cite{bando1}.}. 
Note that the value of $\tan\beta$ does not follow from
the Pendleton-Ross nor from the infrared
quasi--fixed--point behavior of the
Yukawa couplings, because $\rho_r=\rho_{\tau}/\rho_b$ does not have
the infrared behavior like $\rho_t$, as one can see from eq. (5).

\section{Testability of a GYU by $M_t$}
The gross behavior of the Yukawa couplings discussed 
in the previous section 
gives an insight into the GYU physics, and we have seen that
the testability of GYU models and the possibility
to discriminate among them by $M_t$ crucially depend
on the infrared structure of the Yukawa couplings.
In this section,
we include into the evolution of the couplings:
(i) $g_1$ and $g_2$,
(ii) two-loop effects and 
(iii) corrections for the physical masses,
where we neglect
the non-logarithmic threshold corrections such as the
finite corrections coming from the transition from the dimensional
reduction scheme to the 
$\overline{\mbox{MS}}$ scheme \footnote{The 
corrections coming from this transition of
renormalization scheme will be partly taken account
in the next section. Unless it is explicitly stated,
these corrections are not considered below.}.
Then we examine numerically the evolution of the gauge
and Yukawa couplings, according to their RG
 equations  \cite{barger}.
Below $M_{\rm GUT}$ the evolution of couplings is assumed to be
governed by the MSSM. We further assume a unique threshold
$M_{\rm SUSY}$ for all superpartners of the MSSM so that
below $M_{\rm SUSY}$ the SM is the correct effective theory.
The uncertainty in the $M_t$ prediction caused
by these approximations will be discussed  and
estimated when
considering concrete GYU models in the next section.

We  recall that, with a GYU boundary condition at $M_{\rm GUT}$
alone, the value of $\tan\beta$ can not be determined.
Usually, $\tan\beta$ is determined in the Higgs sector, which however
 depends strongly on the supersymmetry breaking terms.
Here we avoid this by using the tau mass $M_{\tau}$ 
as input. (This means that we partly fix the Higgs sector
indirectly.)
That is, assuming that
\be
M_Z \ll M_{t} \ll M_{\rm SUSY}~,
\ee
we require the matching condition at $M_{\rm SUSY}$ \cite{barger},
\be
\alpha_{t}^{\rm SM} 
&=&\alpha_{t}\,\sin^2 \beta~,~
\alpha_{b}^{\rm SM}
~ =~ \alpha_{b}\,\cos^2 \beta~,
~\alpha_{\tau}^{\rm SM}
~=~\alpha_{\tau}\,\cos^2 \beta~,\nn\\
\alpha_{\lambda}&=&
\frac{1}{4}(\frac{3}{5}\alpha_{1}
+\alpha_2)\,\cos^2 2\beta~,
\ee
to be satisfied \footnote{There are MSSM threshold corrections
to this matching condition \cite{hall1,wright1}, 
which will be discussed
later.}, 
where $\alpha_{i}^{\rm SM}~(i=t,b,\tau)$ are
the SM Yukawa couplings and $\alpha_{\lambda}$ is the Higgs coupling.
 This is our definition of $\tan\beta$, and eq. (9)
 fixes $\tan\beta$, because with a given set of the input
parameters \cite{pdg}, 
\be
M_{\tau} &=&1.777 ~\mbox{GeV}~,~M_Z=91.188 ~\mbox{GeV}~,
\ee
with \cite{pokorski1}
\be
\alpha_{\rm EM}^{-1}(M_{Z})&=&127.9
+\frac{8}{9\pi}\,\log\frac{M_t}{M_Z} ~,\nn\\
\sin^{2} \theta_{\rm W}(M_{Z})&=&0.2319
-3.03\times 10^{-5}T-8.4\times 10^{-8}T^2~,\\
T &= &M_t /[\mbox{GeV}] -165~,\nn
\ee
the matching condition (9) and the GYU
boundary condition at $M_{\rm GUT}$ can be satisfied only for a specific
value of $\tan\beta$. Here  $M_{\tau},M_t, M_Z$
are pole masses, and the couplings are defined in the 
$\overline{\mbox{MS}}$ scheme with six flavors.

The translation from a Yukawa coupling
into the corresponding mass follows according to
\be
m_i&=&\frac{1}{\sqrt{2}}g_i(\mu)\,v(\mu)~,~i=t,b,\tau ~~
\mbox{with} ~~v(M_Z)=246.22~\mbox{GeV}~,
\ee
where $m_i(\mu)$'s are the running masses satisfying
the respective evolution equation of two-loop order.
The pole masses can be calculated from the
running ones, of course. For the top mass, we use \cite{barger,hall1}
\be
M_{t} &=&m_{t}(M_t)\,[\,1+
\frac{4}{3}\frac{\alpha_3(M_t)}{\pi}+
10.95\,(\frac{\alpha_3(M_t)}{\pi})^2+k_t 
\frac{\alpha_t(M_t)}{\pi}\,]~,
\ee
where  $k_t \simeq -0.3$ for the range of parameters
we are concerned with in this paper \cite{hall1}.
Note that both sides of eq. (13) contains $M_t$ so that
$M_t$ is defined only implicitly.
Therefore, its determination requires an iteration method.
As for the tau and bottom masses, we assume that
$m_{\tau}(\mu)$ and $m_b(\mu)$ for $\mu \leq M_Z$
satisfy the evolution equation governed by
the $SU(3)_{\rm C}\times U(1)_{\rm EM}$ theory 
with five flavors and use
\be
M_{b}&=&m_b(M_b)\,[\,1+
\frac{4}{3}\frac{\alpha_{3(5{\rm f})}(M_b)}{\pi}+
12.4\,(\frac{\alpha_{3(5{\rm f})}(M_b)}{\pi})^2\,]~,\nn\\
M_{\tau}&=&m_{\tau}(M_{\tau})\,[\,1+
\frac{\alpha_{\rm EM (5f)}(M_{\tau})}{\pi}\,]~,
\ee
where the experimental value of $m_b(M_b)$ is
$(4.1-4.5)$ GeV \cite{pdg}.
The couplings with five flavors entered in eq. (14)
$\alpha_{3(5{\rm f})}$ and $\alpha_{\rm EM (5f)}$
are related to $\alpha_{3}$ and $\alpha_{\rm EM}$ by
\be
\alpha_{3(5{\rm f})}^{-1}(M_Z) &= &\alpha_{3}^{-1}(M_Z)
-\frac{1}{3\pi}\,\ln \frac{M_t}{M_Z} ~,\nn\\
\alpha_{\rm EM (5f)}^{-1}(M_Z) &= & \alpha_{\rm EM}^{-1}(M_Z)-
\frac{8}{9\pi}\,\ln \frac{M_t}{M_Z}~.
\ee
Using the input values given in eqs. (10) and (11), we find
\be
m_{\tau}(M_{\tau})&=&1.771~\mbox{GeV}~,
m_{\tau}(M_{Z})=1.746~\mbox{GeV}~,
\alpha_{\rm EM (5f)}^{-1}(M_{\tau})=133.7~,
\ee
and from eq. (12) we 
obtain 
\be
\alpha_{\tau}^{\rm SM}(M_Z)&=&\frac{g_{\tau}^{2}}{4\pi}
=8.005\times 10^{-6}~,
\ee
which we use as an input parameter instead of $M_{\tau}$.

\begin{figure}
           \epsfxsize= 11 cm   
           \centerline{\epsffile{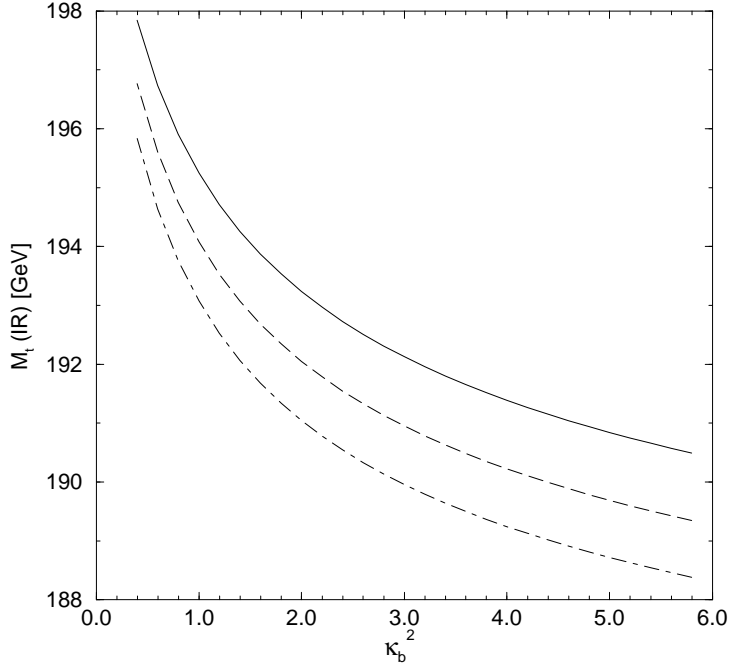}}
        \caption{ $M_t$(IR)
as a function  of $\kappa_{b}^{2}$ for 
 $M_{\rm SUSY} = 1$ TeV (solid),
$500$ GeV (dashed) and $300$ GeV (dot-dashed).}
        \label{fig:1}
        \end{figure}

With these assumptions specified above, we compute
the infrared quasi--fixed--point values 
of the top quark mass $M_t$(IR) \cite{hill1,bardeen}
as  function  of $\kappa_{b}^{2} (=\kappa_{\tau}^{2})$ 
for three different
 $M_{\rm SUSY}$'s, where we fix $\kappa_{t}^{2}$ at $6$. 
(Since
$\alpha_{\rm GUT}\simeq 0.04$,
$\kappa_{t}^{2}=6$ means
that $g_t (M_{\rm GUT}) \simeq 1.7$. 
The values of  $M_t$(IR)
will be increased by  $\sim +1$ GeV if we use
$\kappa_{t}^{2}=8$ ($g_t (M_{\rm GUT}) \simeq 2$.)
This is shown in fig.
1, where the solid, dashed and dot-dashed lines correspond to
$M_t$(IR) with $M_{\rm SUSY}=1000,~500$ and $300$ GeV.
If the predicted values of
$M_t$ are sufficiently below $M_t$(IR), there will be a chance to
discriminate different boundary conditions
of the Yukawa couplings at $M_{\rm GUT}$ and hence
to distinguish different GYU models.
Fig. 1 also shows that a $M_t$  
below $\sim 188$ GeV cannot be solely understood as a consequence
of the infrared quasi--fixed--point behavior of the Yukawa 
couplings in the MSSM.

\begin{figure}
           \epsfxsize= 11 cm   
           \centerline{\epsffile{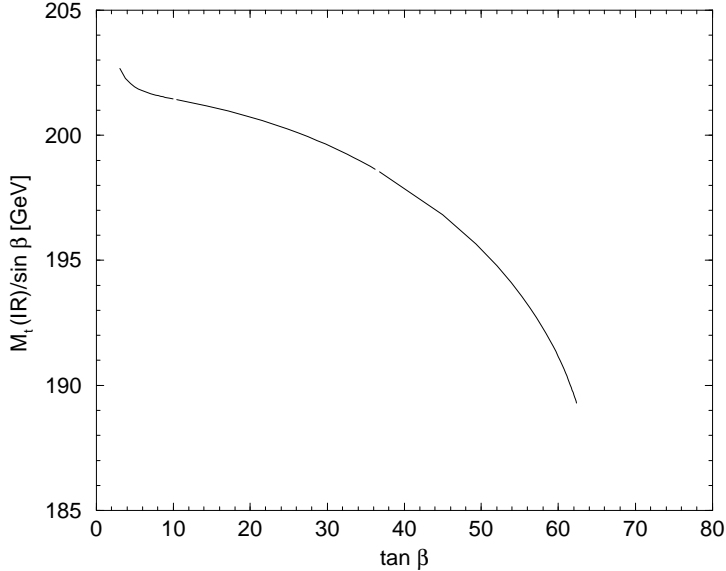}}
        \caption{ $M_t$(IR)$/\sin\beta$
as a function  of $\tan\beta$ for 
 $M_{\rm SUSY} = 500$ GeV.}
        \label{fig:2}
        \end{figure}

Note that the values for $\tan\beta$ in the parameter range
in fig. 1 are large ($\gsim 40$).
In this regime, the ratio $M_t(\mbox{IR})/\sin\beta$
depends on
$\tan\beta$ as one can see from fig. 2.
We find that 
for $7 \lsim \tan\beta \lsim 40$,
\be
M_t(\mbox{IR})/\sin\beta &\simeq &  
\left\{
\begin{array}{l} (203-199)~\mbox{GeV} \\
(201-197)  ~~\mbox{GeV}\\
(199-195) ~~\mbox{GeV}
 \end{array} \right.~~\mbox{for}~~M_{\rm SUSY}~=~
\left\{
\begin{array}{l} 1~\mbox{TeV} \\
500 ~~\mbox{GeV}\\ 300 ~~\mbox{GeV}
 \end{array} \right.~,\nn
\ee
while for $\tan\beta \gsim 40$, $M_t(\mbox{IR})$ 
for $M_{\rm SUSY}=500$ GeV may be approximated
as 
\be
M_t(\mbox{IR}) &\simeq&
195-0.3\,\Delta t-0.01\,(\Delta t)^2~,\\
\Delta t &=& \tan\beta-50~.\nn
\ee
The result above is consistent with
the previous one, $M_t/\sin\beta 
\simeq (190-210)$ GeV \cite{bardeen}.
The $\tan\beta$ dependence of  $M_t/\sin\beta$
for large $\tan\beta$
appears, because $\alpha_b$ and $\alpha_{\tau}$ nontrivially
contribute to the evolution of $\alpha_t.$

\begin{figure}
           \epsfxsize= 11 cm   
           \centerline{\epsffile{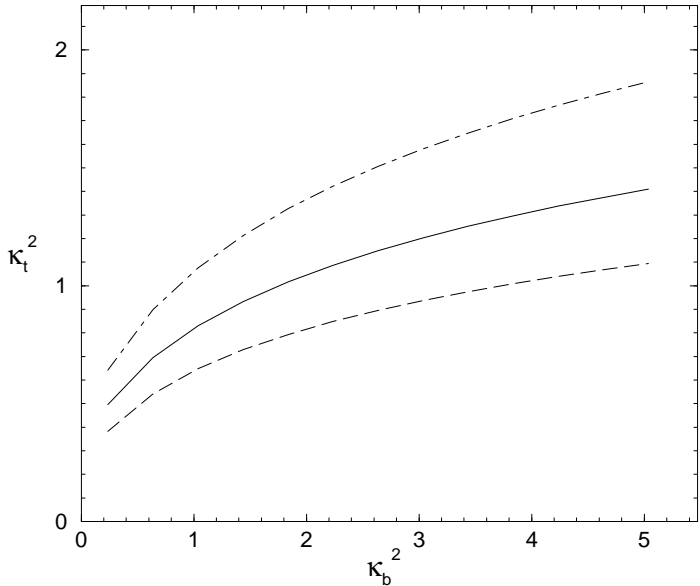}}
        \caption{The equiv-$M_t$ lines for 
 $M_{t} = 182$ (dot-dashed)  ,
$179$ (solid) and $176$ GeV (dashed)
with $M_{\rm SUSY}=500$ GeV.}
        \label{fig:3}
        \end{figure}
Next we would like to come to
the equiv-$M_t$ lines in the $\kappa_{t}^{2}-\kappa_{b}^{2}$
plane 
\footnote{We assume that 
$\kappa_{b}^{2}=\kappa_{\tau}^{2}$.}. 
If one concentrates
only on the $M_t$ prediction from GYUs and discard others,
there have to exist
such lines, because $M_t$ as a function of
$\kappa_{t}^{2}$ and $\kappa_{b}^{2}$ defines a smooth surface.
Strictly speaking, we should
talk about 
 equiv-$M_t$ surfaces, 
because these equiv-$M_t$ lines exist for a given $M_{\rm SUSY}$
which can be seen as a parameter, too.
In fig. 3 we show such lines for $M_{t}=182, 179, 176$ GeV,
where we have fixed $M_{\rm SUSY}$ at $500$ GeV.
Fig. 3 shows that GYU models with $\kappa_b \lsim 1$ can be
better distinguished.
It is clear that to discriminate two models on the same line,
we have to look at other predictions, e.g.,
$M_b$ and $\alpha_{3}(M_Z)$.

The matching condition (9)  suffers from the threshold 
corrections coming from the MSSM superpartners:
\be
\alpha_{i}^{\rm SM} \to 
\alpha_{i}^{\rm SM}(1+\Delta_{i}^{\rm SUSY})~,~i=1,2,\dots,\tau~,
\ee
It was shown that these threshold effects to
the  gauge couplings can
be effectively parametrized by just one energy  scale 
\cite{langacker1}. 
Accordingly, we can identify our $M_{\rm SUSY}$ with that defined
in ref.\cite{langacker1}.  This ensures that there are no further
one-loop threshold corrections  to $\alpha_3(M_Z)$ when we
calculate it as a function of  $\alpha_{\rm EM}(M_Z)$ and
$\sin^2\theta_W(M_Z)$.

The same scale $M_{\rm SUSY}$
does not describe  threshold corrections to the Yukawa
couplings,
and they could cause large corrections to the fermion mass
 prediction \cite{hall1,wright1} \footnote{It is, of course
possible, to compute the MSSM correction to $M_t$ directly, i.e., 
without constructing an effective theory below $M_{\rm SUSY}$.
In this approach, too,  large corrections have  been reported
\cite{polonsky1}. In the present paper, evidently,  we are following
the effective theory approach as 
e.g. refs. \cite{hall1,wright1}.}.
For $m_b$, for instance, the correction  can be as large as 50\%
for very large values  of $\tan\beta$,
especially in models with radiative 
gauge symmetry breaking and with supersymmetry softly broken by 
the universal breaking terms. As we will see
when discussing concrete $SU(5)$ GYU models in the next
section,  these
models predict (with these corrections suppressed) values
for the bottom quark mass that are 
rather close to the experimentally allowed region 
so that there is room only for small corrections.
Consequently, GYU models in which
$SU(2) \times U(1)$ gauge 
symmetry is broken radiatively favor
non-universal soft breaking terms \cite{borzumati1}.
It is interesting to note that
the consistency of the GYU hypothesis
is closely related to the fine structure of supersymmetry breaking
and also to the Higgs sector, because
these superpartner corrections  to $m_b$ can be kept small
for appropriate 
supersymmetric spectrum characterized by very heavy squarks 
and/or small  $\mu_H$ describing the mixing of the two 
Higgs doublets in the superpotential 
\footnote{The solution with small $\mu_H$ 
is favored by the experimental data and cosmological constraints
\cite{borzumati1}. The sign of this correction 
is determined by the relative sign of 
$\mu_H$ and the gluino mass parameter, $M_3$, and is correlated 
with the chargino exchange contribution 
to the $b \to s \gamma$ decay \cite{hall1}. 
The later has the same sign as the Standard Model and the charged 
Higgs contributions when the supersymmetric corrections to $m_b$ are 
negative.}.

To get an idea about the magnitude of the correction, 
let us consider
the case that all the superpartners 
have the same mass $M_{\rm SUSY}$ with
$M_{\rm SUSY} >> \mu_H$ and $\tan\beta \gsim 50$, and vary
$\mu_H/M_{\rm SUSY}$ from $-0.2$ to $0.15$.
We quote $\Delta$'s at $\mu=M_{\rm SUSY}$
from ref. \cite{wright1}:
 \be
2\pi \Delta_t &\simeq& -\frac{4}{3}\alpha_3
-\frac{1}{8}\alpha_b
~,\nn \\ 
2\pi \Delta_b &\simeq&
-\frac{4}{3}\alpha_3+\frac{1}{4}\alpha_b F_2(M_{\rm SUSY}^2,
M_{t}^{2})+ ( -\frac{4}{3}\alpha_3+\frac{1}{2}\alpha_2
-\frac{1}{2}\alpha_t)~\frac{\mu_H}{M_{\rm SUSY}}\tan\beta ~~,\nn\\
2\pi \Delta_{\tau} &\simeq& -\frac{1}{4}\alpha_2
-\frac{1}{8}\alpha_{\tau}+\frac{1}{4}\alpha_2 
\frac{\mu_H}{M_{\rm SUSY}}\tan\beta ~,\nn
\ee
where 
\be
F_2(M_{\rm SUSY}^2, M_{t}^{2}) &\simeq&
-\frac{1}{2} +\frac{M_{t}^{2}}{M_{\rm SUSY}^{2}} 
+  \frac{M_{t}^{4}}{M_{\rm SUSY}^{4}} 
\ln(\frac{M_{t}^{2}}{M_{\rm SUSY}^{2}})~\mbox{for}~
M_t \ll M_{\rm SUSY}~,\nn
 \ee
and terms proportional to $\cos\beta, \cot\beta$ etc.
are suppressed. The result is presented in  table 1,
where it is assumed: $M_{\rm SUSY}=500$ GeV and
a GYU boundary condition $\alpha_3=\alpha_2=\alpha_1=
\alpha_t=\alpha_b=\alpha_{\tau}
=\alpha_{\rm GUT}$.

\begin{center}
\begin{tabular}{|c|c|c|c|c|c|c|}
\hline
$\mu_H/M_{\rm SUSY}$ &
$M_t$[GeV] &  $m_b(M_b)$ [GeV]
\\ \hline
$-0.2 $  &$ 175.4 $  & $5.02 $ \\ \hline
 $-0.1 $  &$ 175.4 $  & $ 4.80$ \\ \hline
 $0.1 $  &$ 175.4 $  & $ 4.31$  \\ \hline
 $0.15 $  &$ 175.4 $  & $ 4.18$ \\ \hline
\end{tabular}
\end{center}

\begin{center}
{\bf Table 1}. The $\mu_H/M_{\rm SUSY}$ dependence
of the top and bottom quark mass predictions.
\end{center}

\vspace{0.2cm}
\noindent
Without the threshold corrections $\Delta$'s, we obtain
$M_t=177.2   $ GeV and $m_b (M_b)=  4.62$ GeV, and so 
the MSSM correction to the $M_t$ prediction
is $\sim -1$ \% in this case.
Comparing with the 
results of \cite{wright1,polonsky1}, 
this may appear to be underestimated.
Note however that there is a nontrivial interplay among the 
corrections between the $M_t$ and $M_b$ predictions
for a given GYU boundary condition at $M_{\rm GUT}$
and the fixed pole tau mass, which has not been taken into
account in refs. \cite{wright1,polonsky1}. 
Here
 we will not go into details of this MSSM
correction,
and we leave this problem to future work.
In the following discussions, the MSSM threshold correction to
the $M_t$ prediction, which will
become calculable once the superpartner
spectra are known, will be denoted by
\be
\delta^{\rm MSSM} M_t~.
\ee

\section{Comparison of the Gauge-Yukawa 
Unified models based on $SU(5)$}
There are two gauge-Yukawa unified models based 
on $SU(5)$ we would like to consider here;
the first one 
\cite{kubo2}
is an asymptotically-free unified theory (AFUT), and the second one 
(FUT) \cite{mondragon1}
 is finite to all orders.
Here we would like to address the question
whether it is possible to distinguish
these models experimentally.

\subsection{Asymptotically Free Unified Theory}
The field content is minimal \cite{sakai1}: three
generations of quarks and leptons   are accommodated by 
three chiral supermultiplets in
$\Psi^{I}({\bf 10})$ and $\Phi^{I}(\overline{\bf 5})$,
where $I$ runs over the three generations.
A $\Sigma({\bf 24})$ is used to break $SU(5)$ down to $SU(3)_{\rm C}
\times SU(2)_{\rm L} \times U(1)_{\rm Y}$,  and
$H({\bf 5})$ and $\overline{H}({\overline{\bf 5}})$
 to describe the
two Higgs supermultiplets appropriate for 
electroweak symmetry breaking.
The superpotential of the model is given by \cite{sakai1}
\be
W &=& \frac{g_{t}}{4}\,
\epsilon^{\alpha\beta\gamma\delta\tau}\,
\Psi^{3}_{\alpha\beta}\Psi^{3}_{\gamma\delta}H_{\tau}+
\sqrt{2}g_b\,\overline{H}^{\alpha}
\Psi^{3}_{\alpha\beta}\Phi^{3 \beta}+
\frac{g_{\lambda}}{3}\,\Sigma_{\alpha}^{\beta}
\Sigma_{\beta}^{\gamma}\Sigma_{\gamma}^{\alpha}+
g_{f}\,\overline{H}^{\alpha}\Sigma_{\alpha}^{\beta} H_{\beta}\nn\\
& &+ m_{1}\, \Sigma_{\alpha}^{\gamma}\Sigma_{\gamma}^{\alpha}+ 
+m_{2}\,\overline{H}^{\alpha} H_{\alpha}~,
\ee
where $\alpha,\beta,\ldots$ are the $SU(5)$
indices, and we have suppressed the Yukawa couplings of the
first two generations.
In this approximation, there are two Yukawa and two Higgs
couplings. If one applies the principle of reduction of couplings
and require the theory to be asymptotically free, one finds that
$\kappa_t$ and $\kappa_b=\kappa_{\tau}$ are strongly 
constrained \cite{kubo2}.
In the one-loop approximation, they can be written as
\be
\kappa_{t,b}^{2} &=&\sum_{m,n=0}^{\infty}
\,r_{t,b}^{(m,n)}\,
[\tilde{\alpha}_{\lambda}]^{m}\,[\tilde{\alpha}_{f}]^{n}~,
\ee
where
\be
\tilde{\alpha}_{i} &\equiv &\alpha_i /\alpha~,~i=\lambda,f~,
\ee
and the first expansion coefficients $r_{t,b}^{(m,n)}$
are given in table 2.

\vspace{1cm}
\begin{center}
\begin{tabular}{|c|c|c||c|c|c|}
\hline
$m,n$  &$r_{t}^{(m,n)}$ &
$r_{b}^{(m,n)}$ &
 $m,n$  &  $r_{t}^{(m,n)}$ 
 & $r_{b}^{(m,n)}$
\\ \hline
$0,0$ & $89/65$ &$63/65$ &  $1,2$ & $
-0.0114$ 
 & $-0.0111$ \\ \hline
$1,0$  & $ 0$ &$0$ & $0,3$ & $ -0.0225$  
 & $-0.0226$  \\ \hline
$0,1$ & $-0.2581 $  &$-0.2131$ & $4,0$ & $0$ 
 & $0$\\ \hline
$2,0$ & $0$  &$0$ & $3,1$ & $ -0.0012$  
 & $-0.0012$ \\ \hline
$1,1$ & $-0.0205$  &$-0.0184$ &  $2,2 $& $-0.0034  $
 & $ -0.0035 $ \\ \hline
$0,2$ & $ -0.0549$  &$ -0.0501$ & $1,3 $ &  
$ -0.0074 $ 
 & $  -0.0079$\\ \hline
$3,0$ & 
$0$ & $0$  & $0,4$ &   $-0.0114$  
 & $  -0.0126$\\ \hline
$2,1$ & $ -0.0043$   
 & $-0.0041$  & \vdots &   \vdots
 & \vdots \\ \hline
\end{tabular}
\end{center}

\begin{center}
{\bf Table 2}. The expansion coefficients $r_{t,b}^{(m,n)}$
for $m+n \leq 4$.
\end{center}

\noindent
The coupling $\tilde{\alpha}_{\lambda}$ is allowed to vary from 
$0$ to $15/7$, while $\tilde{\alpha}_{f}$ may vary from $0$ to a maximum 
$\tilde{\alpha}_{\rm fmax}$ which depends
on $\tilde{\alpha}_{\lambda}$. 
For small $\tilde{\alpha}_{\lambda}$, it is given by
\be
\tilde{\alpha}_{\rm fmax} &=&
560/521 - 0.1313\ldots  \tilde{\alpha}_{\lambda} - 
  0.0212\ldots [\tilde{\alpha}_{\lambda}]^2 
- 0.0058\ldots [\tilde{\alpha}_{\lambda}]^3 \nn\\
& &- 0.0019\ldots [\tilde{\alpha}_{\lambda}]^4
+O([\tilde{\alpha}_{\lambda}]^5)~,
\ee
and, it vanishes at $\tilde{\alpha}_{\lambda}=15/7$.
Therefore, each equation of (22) describes a two-dimensional
surface with boundary.
As we can see from eq. (22), along with the
$r_{t,b}^{(m,n)}$'s given in table 2, the 
$\tilde{\alpha}_{\lambda}$-dependence
of $\kappa$'s are rather weak, so we show in table 3 the
predictions
for different values of $\tilde{\alpha}_{f}$, where 
$\tilde{\alpha}_{\lambda}$ is fixed  at zero.
($M_{\rm SUSY}=500$ GeV)

\vspace{1cm}

\begin{tabular}{|c|c|c|c|c|c|c|c|}
\hline
$\tilde{\alpha}_f$   & $\kappa_{t}^{2}$ & $\kappa_{b}^{2}$&
$\alpha_{3}(M_Z)$ &
$\tan \beta$  &  $M_{\rm GUT}$ [GeV] 
 & $m_b (M_{b}) $ [GeV]& $M_{t}$ [GeV]
\\ \hline
$0.2$ & $1.315$& $0.925$  & $0.122 $  & $52.2 $ & 
 $1.74\times 10^{16} $ & $4.60 $  &$180.9 $  \\ \hline
 $0.6$ & $1.187$& $0.816$ &  $0.121 $  & $51.1 $ & 
 $1.73\times 10^{16} $ & $4.63 $  &$179.8 $  \\ \hline
 $1.0$ & $1.001$& $0.642$ & $0.121 $  & $49.0 $ & 
 $1.71\times 10^{16} $ & $4.69 $  &$179.1 $   \\ \hline 
$1.075$ & $0.972$& $0.572$ & $0.121 $  & $47.9 $ & 
 $1.71\times 10^{16} $ & $4.72 $  &$177.8 $   \\ \hline
\end{tabular}

\begin{center}
{\bf Table 3}. The predictions of AFUT
for  $M_{\rm SUSY}
=500$ GeV (before the threshold corrections from
the superheavy particles).
\end{center}

\noindent
>From table 3 we see that the predicted values for 
$M_t$ are well below the infrared 
values \footnote{For $\kappa_{b}^{2}=0.64$ and
$M_{\rm SUSY}=500$ GeV, for instance, the $M_t$(IR)
is $\simeq 195$ GeV.} and 
the width of the $M_t$ prediction on the asymptotically
free surface is at most $4$ GeV. But this
uncertainty can be shrunk as we will do now.

We  turn our discussion to proton decay
to further constrain the parameter space of AFUT.
First we recall that if one includes
the threshold effects of superheavy particles \cite{threshold},
the GUT scale $M_{\rm GUT}$ at which $\alpha_1$ and $\alpha_2$
are supposed to meet is related to
$M_H$, the mass of the superheavy
$SU(3)_C $-triplet  Higgs supermultiplets contained 
in $H$ and $\overline{H} $.
To see this, we write the one-loop relation
\be
\alpha_{2}^{-1}(\mu) &=&
\alpha^{-1}(\Lambda)+\frac{1}{2\pi}\,(\,
\ln\frac{\Lambda}{\mu}-6\ln\frac{\Lambda}{M_V}+
2\ln\frac{\Lambda}{M_{\Sigma}}\,)~,~\nn\\
\alpha_{1}^{-1}(\mu) &=&
\alpha^{-1}(\Lambda)+\frac{1}{2\pi}\,(\,
\frac{33}{5}\ln\frac{\Lambda}{\mu}-10\ln\frac{\Lambda}{M_V}+
\frac{2}{5}\ln\frac{\Lambda}{M_{H}}\,)~,
\ee
where $M_{\Sigma}$ and $M_{V}$ stand for the masses of the
superheavy Higgs supermultiplets contained in ${\bf 24}$ and
the superheavy gauge supermultiplets, and
$\mu > M_{\rm SUSY}$ and $\Lambda > M_V ,M_H ,M_{\Sigma}$.
Then from 
$\alpha_{1}^{-1}(M_{\rm GUT}) =\alpha_{2}^{-1}(M_{\rm GUT})$,
we find that
\be
M_{\rm GUT} &=& [M_V]^{5/7}\, [M_H]^{-1/14}\, 
[M_{\Sigma}]^{5/14}~.
\ee
Using the tree-level mass relations,
\be
\frac{M_V}{M_{\Sigma}} &=& 2\sqrt{2}\,\frac{g}{g_{\lambda}}~,~
\frac{M_H}{M_{\Sigma}} ~=~ 2\,\frac{g_f}{g_{\lambda}}~,
\ee
which follows from the assumption that 
the mass parameter $m_2$ in the superpotential is fine tuned so that the
 $SU(2)_{L}$-doublet Higgs supermultiplets remain light,
we can rewrite eq. (26) as
\be
M_H &=&[\tilde{\alpha}_{f}]^{15/28}\,
[\tilde{\alpha}_{\lambda}]^{-5/28}\,M_{\rm GUT}~.
\ee
As known \cite{d5}, $M_H$ controls the nucleon decay which 
is mediated by dimension five operators, and
non-observation of the nucleon decay requires
$M_H \gsim 10^{17}$ GeV for $\tan\beta\simeq 50$ \cite{hisano1}.
Since $M_{\rm GUT} \simeq 1.7\times 10^{16}$ GeV and
$\tilde{\alpha}_{f} \lsim 1.1 $ as 
one can see from eq. (22) and table 2, the value of 
$\tilde{\alpha}_{\lambda}$ has to be less than
$\sim 4.4 \times 10^{-5}$. Therefore, the reduction solutions that
are consistent with the nucleon decay constraint are very close to
 the boundary of the asymptotically
surface, i.e., $\tilde{\alpha}_{f}$ has to be very close
to $1.075$ (see table 3). Comparing $\kappa$'s for
large $\tilde{\alpha}_{f}$ in table 3, we assume in the following
discussions that for AFUT
\be
\kappa_{t}^{2} &=&1.0 ~~\mbox{and}~~\kappa_{b}^{2} ~=~0.64 ~,
\ee
with an uncertainty of $5~(10)$ \%  for 
$\kappa_{t}^{2}~(\kappa_{b}^{2})$.

\begin{figure}
           \epsfxsize= 11 cm   
           \centerline{\epsffile{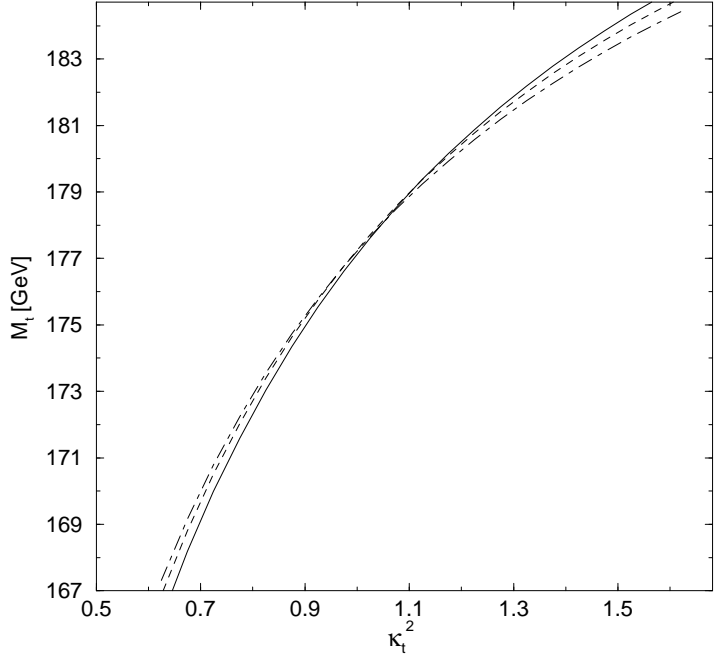}}
        \caption{$M_t$ against $\kappa_{t}^{2}$  
($\kappa_{b}^{2}=0.64$)  
with $M_{\rm SUSY}$ 
=$1000 (solid), 500 (dashed)$ and $300$ (dod-dashed) GeV. }
        \label{fig:4}
        \end{figure}
\begin{figure}
           \epsfxsize= 11 cm   
           \centerline{\epsffile{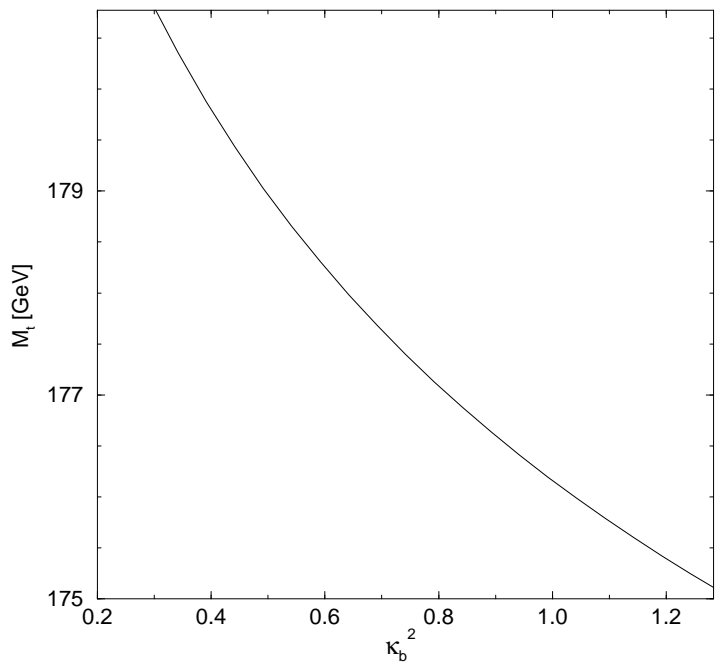}}
        \caption{$M_t$ against $\kappa_{b}^{2}$  
($\kappa_{t}^{2}=1.0$)  
with $M_{\rm SUSY} = 500$ GeV. }
        \label{fig:5}
        \end{figure}

In fig. 4 we show $M_t$ as a function of $\kappa_{t}^{2}$
with $M_{\rm SUSY}=1000, ~500$ and $300$ GeV, where
$\kappa_{b}^{2}$ is fixed at the predicted value $0.64$.
>From this we see that around $\kappa_{t}^{2}=1$ the $M_{t}$
prediction is sensitive against the change
of $\kappa_{t}^{2}$ and  is stable against the change of 
$M_{\rm SUSY}$.
Fig. 5 shows the sensitivity of the $M_t$ prediction
against $\kappa_{b}^{2}$ with $\kappa_{t}^{2}$ fixed at $1.0$
($M_{\rm SUSY}=500$ GeV).
>From these figures we may conclude that accurate measurements on $M_{t}$
with an uncertainty of less than $\sim $few GeV would
exclude or confirm the predicted region in  the
$\kappa_{t}^{2}-\kappa_{b}^{2}$ plane in AFUT
if there would be no theoretical uncertainty.

The threshold effects of the superheavy particles has also
an influence on the coupling 
unification \footnote{It has been 
argued \cite{langacker1,hall2}
that there might 
be gravitationally  induced effects that cause 
 corrections to the coupling unification.
Here we do not consider them, because
it has not been well established how to introduce
gravitational effects into the framework of renormalizable
quantum field theory.},
and we would like to calculate them.
 One finds
in one-loop order that
\be
\alpha_{3}^{-1}(M_{\rm GUT}) &=&
\alpha_{\rm GUT}^{-1}+\frac{1}{2\pi}\,[\,
\frac{6}{7}\ln\frac{M_V}{M_{\Sigma}}-
\frac{9}{7}\ln\frac{M_H}{M_{\Sigma}}\,]\nn\\
&=&
\alpha_{\rm GUT}^{-1}+\frac{1}{2\pi}\,[\,
\frac{3}{14}\ln\tilde{\alpha}_{\lambda}-
\frac{9}{14}\ln\tilde{\alpha}_{f}\,]~,
\ee
where we have used the mass relations (27), and
$\alpha_{\rm GUT}^{-1}=
\alpha_{2}^{-1}(M_{\rm GUT})=\alpha_{1}^{-1}(M_{\rm GUT})$.
Since 
$\tilde{\alpha}_{f} \simeq 1$ and 
$\tilde{\alpha}_{\lambda} \lsim 4.4\times 10^{-5}$ 
because of the proton decay constraint (as discussed above)and 
$\alpha_{\rm GUT}\simeq 0.04$, we obtain
\be
\alpha_{3}(M_{\rm GUT}) &\gsim & 1.014\,\alpha_{\rm GUT}~.
\ee
In table 4 we present the predictions with the
threshold effects
(i.e., $\alpha_{3}(M_{\rm GUT}) = 1.014\,\alpha_{\rm GUT}$), where
we have used $\kappa_{t}^{2}=1.0$ and $\kappa_{b}^{2}=0.64$. 
The numbers in the parenthesis are those obtained without the
threshold effects, i.e.,
$\alpha_{3}(M_{\rm GUT}) = \alpha_{\rm GUT}$.

\vspace{1cm}
\begin{center}
\begin{tabular}{|c|c|c|c|c|}
\hline
$M_{\rm SUSY}$ [TeV]   & 
$\alpha_{3}(M_Z)$ 
 & $m_b (M_{b}) $ [GeV]& $M_{t}$ [GeV]
\\ \hline
$0.5$ & $0.127(0.121)$ & $4.89(4.69) $  &$180.6(177.8) $  \\ \hline
$1$ & $0.125(0.119) $  & $4.87(4.68) $  &$180.5(177.7) $  \\ \hline
\end{tabular}
\end{center}

\begin{center}
{\bf Table 4.} The threshold effects of the superheavy particles.
\end{center}

\noindent
As we can see from table 4, the threshold corrections have
significant effects, especially on 
$\alpha_{3}(M_Z)$  and
  $m_b (M_{b}) $ \footnote{As for $m_b (M_{b}) $, 
we see that only negative corrections
to $m_b (M_{b}) $ of at most $20$ \%
(which we have mentioned in the previous section) are allowed.}.
 The predicted values for $\alpha_{3}(M_Z)$
 are slightly larger than the central experimental value
$0.118\pm 0.006$, but
prefers those obtained from the  electroweak data
and ${\em e}^{+}{\em e}^{-}$ jets experiments \cite{hagiwara1}.
Since $\alpha_{3}(M_Z)$ decreases as $M_{\rm SUSY}$ increases,
AFUT needs a relatively large  $M_{\rm SUSY}$.

Now we come to the final value of $M_t$ for AFUT, and
we collect the uncertainties:
The MSSM  threshold
correction  is denoted
 by $\delta^{\rm MSSM} M_t$
as we have discussed in the previous section (see eq. (20)).
The uncertainties involved in $\kappa_{t}^{2}$
and $\kappa_{b}^{2}$ may lead to another
$\sim \pm 1.4$ GeV (see eq. (29) and figs. 4 and 5).
The threshold effects of the superheavy particles are included
only in the gauge sector, but not in the Yukawa sector.
If we assume that its magnitude is similar to that in the gauge sector
(eq. (31)), it will be $\lsim 2$ \% in $\kappa_{b}^{2}$ and
$\kappa_{t}^{2}$, leading to an uncertainty 
of $\sim \pm 0.4$ GeV in $M_t$.
The finite corrections coming from the conversion
from the dimensional reduction (DR) scheme to
the ordinary $\overline{\mbox{MS}}$  at $M_Z$
may be included in the gauge sector \cite{anton}
($\sim 1\%$ for $\alpha_{3}$ and 
$\sim 0.2\%$ for $\alpha_{2}$).
As for the Yukawa sector, we assume that it is similar to that 
for  $\alpha_{3}$,  which gives rise to an uncertainty
of $\sim\pm 1$ GeV in $M_t$. 
>From these
considerations, we finally obtain
\be
M_t &=& (181+\delta^{\rm MSSM} M_t \pm 3) ~~\mbox{GeV}
\ee
for AFUT.

\subsection{Finite Unified Theory} 
>From the classification of 
finite theories given in ref.
\cite{finite2}, one finds that 
using $SU(5)$ as gauge group there
exist only two candidate models which can accommodate three 
generations. 
But only one of them contains a ${\bf 24}$ which can be used
for the spontaneous symmetry breaking SSB of $SU(5)$ down
to $SU(3)\times SU(2) \times U(1)$. For the other one,
one has to incorporate another way, such as the Wilson flux
breaking to achieve the desired SSB of $SU(5)$.
Here we focus our attention on the first model.

The field content is: three 
($\overline{{\bf 5}}+{\bf 10}$)'s
for three generations,
four pairs of (${\bf 5}+\overline{{\bf 5}}$)-Higgses 
and one ${\bf 24}$.
To ensure finiteness 
to all orders \cite{sibold1}, we have to find 
isolated, non-degenerate solutions to
the reduction equations \cite{zimmermann1} of the model,
which are consistent with the 
one-loop finiteness conditions.
In the previous studies of refs. \cite{finite3}, however,
no attempt was made to find isolated, non-degenerate
solutions, but rather the opposite; the freedom
offered by the degeneracy has been used 
in order to make specific
 ans{\" a}tze that could lead to phenomenologically acceptable
predictions (see for 
another attempt ref. \cite{finite4}
in which the dimensional regularization plays the 
fundamental r\^ole). Here we would like
to follow the treatment of ref. \cite{mondragon1}
in which an isolated, non-degenerate
reduction solution exists thanks to certain discrete  symmetries
in the superpotential.
The solution corresponds to
the Yukawa matrices
without intergenerational mixing (which is reasonable 
 as a first approximation), and yields
in the one-loop approximation
\be
\kappa_1 &=&\kappa_2~=~\kappa_3,\nn\\
\kappa_t &=&\kappa_c~=~\kappa_u~=~\sqrt{8/5}~,\nn\\
\kappa_b &=& \kappa_s ~=~\kappa_d ~=~
\kappa_{\tau} ~=~\kappa_{\mu} ~=~\kappa_e
~=~\sqrt{6/5}~. \nn
\ee
At first sight, this GYU boundary condition seems to lead 
to unacceptable predictions of the fermion masses.
But this is not the case, because each generation has 
an own pair of ($\overline{{\bf 5}}+{\bf 5}$)-Higgses:
We may use the fact that mass terms
do not influence the $\beta$-functions in a certain
class of renormalization schemes, and introduce
appropriate mass terms that permit us to perform a rotation in the Higgs
sector such that only one pair of Higgs doublets, coupled to
the third family, remains light and acquires a
non-vanishing VEV (in a similar way to what was done by
Le\'on et al.~\cite{leon1}). 
Note that the effective coupling of the Higgs doublets
to the first family after
the rotation is very small avoiding in this way a potential problem
with the proton lifetime \cite{proton1}.
Thus, effectively,
we have at low energies the MSSM with
only one pair of Higgs doublets.
In other words, the effective GYU boundary condition
under this assumption becomes
\be
\kappa_1 &=&\kappa_2~=~\kappa_3,
\kappa_t ~=~\sqrt{8/5}~,
\kappa_b  ~=~
\kappa_{\tau}~=~\sqrt{6/5}~,
\ee
where the Yukawa couplings of the first two generations
should  be regarded as free parameters.
The predictions of $M_t$ and $m_b(M_b)$ for various $M_{\rm SUSY}$
are given in table 5.

\vspace{1cm}

\begin{center}
\begin{tabular}{|c|c|c|c|c|c|}
\hline
$M_{\rm SUSY}$ [GeV]   &$\alpha_{3}(M_Z)$ &
$\tan \beta$  &  $M_{\rm GUT}$ [GeV] 
 & $m_b (M_{b}) $ [GeV]& $M_{t}$ [GeV]
\\ \hline
$300$ & $0.123 $  &$54.2 $  & $2.06\times 10^{16}$
 & $4.53$  & 182.2\\ \hline
$500$ & $0.122 $  &$54.3 $  & $1.75\times 10^{16}$
 & $4.53$  & 182.6 \\ \hline
$10^3$ & $0.119 $  &$54.4 $  & $1.41\times 10^{16}$
 & $4.53$  & 183.0 \\ \hline
\end{tabular}
\end{center}

\begin{center}
{\bf Table 5}. The predictions 
for different $M_{\rm SUSY}$ for FUT.
\end{center}

\noindent
Similar to the case of AFUT, only negative MSSM corrections
of at most $\sim 10$ \% to $m_b(M_b)$ is allowed, implying that FUT also 
favors  non-universal soft symmetry breaking terms.
The predicted $M_t$ values are well below the infrared value,
for instance
$194$ GeV for $M_{\rm SUSY}=500$ GeV (see fig. 1), 
so that the $M_t$ prediction
must be sensitive against the change of $\kappa_{t}^{2}$
as well as $\kappa_{b}^{2}$. This is shown in figs. 6 and 7.
For fig. 6 we have fixed $\kappa_{b}^{2}$ at $1.2$ and used
$M_{\rm SUSY}=1000$ (solid), $500$ (dashed) and $300$ (dot-dashed) GeV,
while we have fixed  $\kappa_{t}^{2}$ at $1.6$ and used
$M_{\rm SUSY}=500$ in fig. 7.

\begin{figure}
           \epsfxsize= 11 cm   
           \centerline{\epsffile{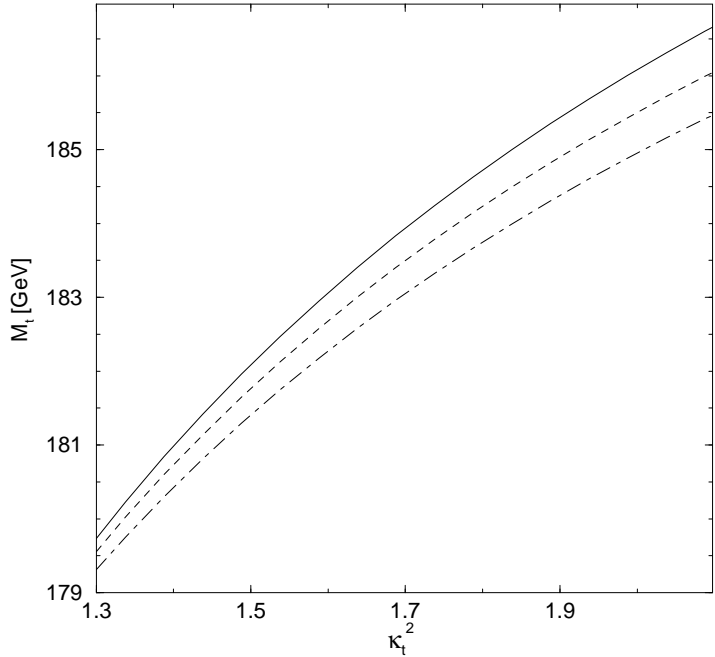}}
        \caption{$M_t$ versus $\kappa_{t}^2$
   for $M_{\rm SUSY}=1000$ (solid),
$500$ (dashed) and $300$ GeV (dot-dashed) 
with $\kappa_{b}^{2}=1.2$ fixed.}
        \label{fig:6}
        \end{figure}

\begin{figure}
           \epsfxsize= 11 cm   
           \centerline{\epsffile{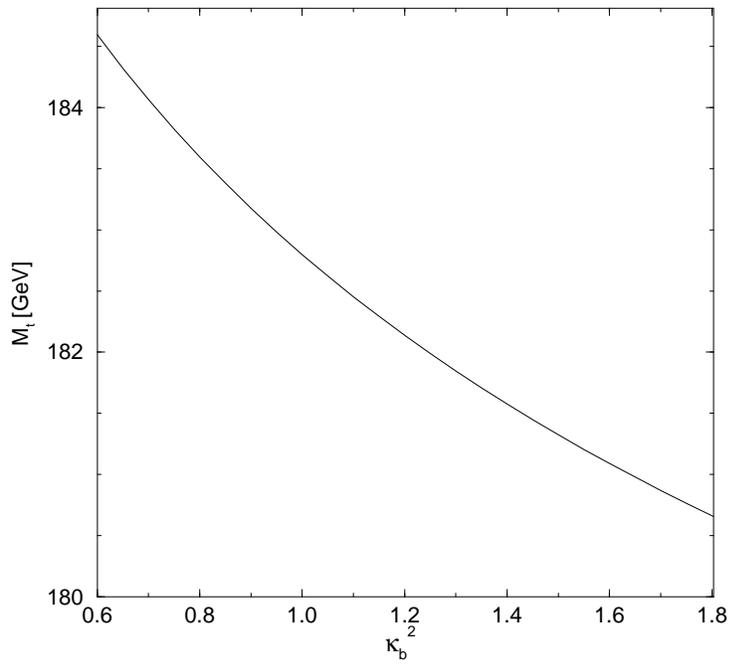}}
        \caption{$M_t$ versus $\kappa_{b}^2$
   for $M_{\rm SUSY}=500$ with $\kappa_{t}^{2}=1.6$ fixed.}
        \label{fig:7}
        \end{figure}

The nice feature of FUT is that there is no finite range in the
$\kappa_{t}^{2}-\kappa_{b}^{2}$ plane; it predicts a point.
But the structure of the threshold effects from the superheavy 
particles is involved in this case, compared to AFUT.
They are not arbitrary and probably determinable to a certain
extent, because the mixing of the superheavy Higgses
is strongly dictated by the fermion mass matrix of the MSSM.
To bring these threshold effects under control is
 beyond the scope of the present
paper, and we would like to leave it to future work.
Here we assume that the magnitude of these effects is
the same as that for AFUT and so 
$\sim \pm 3$  GeV in $M_t$ (see table 4).
The superheavy particle threshold effects in the Yukawa sector
can be estimated from figs. 6 and 7 if we assume that 
its magnitude is also the same as that of AFUT, that is, 
 $2$ \% in $\kappa_{t}^{2}$ and $\kappa_{b}^{2}$.
This gives an uncertainty of $\pm 0.4$  GeV in $M_t$.
The MSSM  threshold
correction 
is  denoted $\delta^{\rm MSSM} M_t$
as for the case of AFUT (see also eq. (20)).
Thus, including all the uncertainties we have discussed above, 
we may conclude that 
\be
M_t &=&(183+\delta^{\rm MSSM} M_t+\pm 5) ~~\mbox{GeV}~,
\ee
for FUT,
where the finite corrections coming from the conversion
from the DR scheme to
the ordinary $\overline{\mbox{MS}}$ 
in the gauge sector \cite{anton} are included, and
those in the Yukawa sector are included as an uncertainty
of $\sim\pm 1$ GeV.

Comparing this with the $M_t$ prediction of AFUT
given in eq. (32), we see that at present
the two existing GYU models based on $SU(5)$
cannot be theoretically distinguished by $M_t$.
To distinguish them, it is therefore important to 
reduce the uncertainties in $M_t$.
Also important is the structure of the supersymmetry breaking,
because, as we see from tables 4 and 5, the two models predict
different $m_b$. Since the accurate prediction
on $m_b$ depends strongly on the soft 
supersymmetry breaking terms, we have to clarify
this subject more in detail in order to distinguish
between AFUT and FUT, which will be our future work.

As a last remark we would like to mention that
even the soft supersymmetry breaking terms can
be controlled by the reduction of couplings and
finiteness \cite{jack}.

\section{Conclusion}
 The electroweak and strong interactions
can be unified in GUTs
\cite{pati1,georgi2,fritzsch1}, thereby relating 
the apparently independent
gauge couplings of the SM.
The observed hierarchy of these couplings, 
$\alpha_1 < \alpha_2 < \alpha_3$,
can be understood if one assumes that
the unifying gauge symmetry is broken  at a 
$M_{\rm GUT}$ which is much larger than the
electroweak scale \cite{gqw}.
The top-bottom mass hierarchy at low energies
could be explained to a certain extent if one assumes the existence of a  GYU
at $M_{\rm GUT}$ \cite{mondragon1}
\cite{kubo2}--\cite{kubo33}. Of course, the observed top-bottom
hierarchy, $M_t /m_b (M_b) \simeq (37-47)$,
is not a proof for a GYU, but it may indicate a unification
that goes beyond the usual grand unification.

We have seen that different GYU
 models could be  distinguished and tested by
a precise measurement of $M_t$ if
 the models are not on the equiv-$M_t$ lines that are very 
close to each other and
the predicted $M_t$'s are well separated
from   the infrared value. 
We, therefore, have analyzed the infrared
quasi--fixed--point behavior of the $M_t$ prediction
in some detail.
We have also seen that the infrared value, $M_t$(IR),
depends on $\tan\beta$ and
its lowest value is $\sim 188$ GeV.
Comparing this with the experimental value \cite{top},
$M_t=(180\pm 12)$ GeV \cite{top}, we may conclude
that the present data on $M_t$ cannot be 
satisfyingly explained solely
from the infrared quasi--fixed--point behavior
of the top Yukawa coupling.
Two GYU models on the same equiv-$M_t$ line
predict in general different values for
$M_b$ and  $\alpha_{3}(M_Z)$, and so
their precise knowledge
will allow us to further shrink the allowed
range of the GYU boundary conditions.

 The main conclusion of our calculations
in AFUT and FUT is that they predict
$M_t =(181+\delta^{\rm MSSM} M_t\pm 3)$ GeV
 and $(183+\delta^{\rm MSSM} M_t\pm 5)$ GeV,
respectively, 
where $\delta^{\rm MSSM} M_t$ stands  for the MSSM threshold
correction.
 We found it is $\sim -2$ GeV for the case that
all the superpartners have the same mass $M_{\rm SUSUY}$
and $\mu_H/M_{\rm SUSUY} \ll 1$.
These predictions are consistent with
the  present experimental data. 
Clearly, to exclude or verify these GYU models,
 the experimental as well as theoretical uncertainties
have to be further reduced.
One of the largest theoretical uncertainties 
 for FUT results
from the not-yet-calculated threshold effects 
of the superheavy particles.
Since the structure of  the superheavy 
particles in FUT is basically fixed,
 it will be possible to
bring these threshold effects under control,
which will  reduce the uncertainty of 
the $M_t$ prediction ($5$ GeV) down to $\sim  2$ GeV.
 
Here we have been regarding $\delta^{\rm MSSM} M_t$ 
as unknown because we have no
sufficient information on the superpartner spectra.
Recently, however, we have found that the idea
of reduction of couplings can be applied to 
dimensionfull parameters too. As a result, 
it becomes possible to predict the superpartner spectra 
to a certain extent and then to calculate $\delta^{\rm MSSM} M_t$.

\end{document}